\documentclass[11pt]{article}
\usepackage{texdraw}
\usepackage[english]{babel}
\begin{document}


\title {The R.I.~Pimenov unified gravitation and electromagnetism field theory as
semi-Riemannian geometry    }         

\author {N.A.~Gromov   \\
 Department of Mahematics, Komi Science Center UrD, RAS          \\
Kommunisticheskaya st. 24, Syktyvkar,167982, Russia \\
E-mail: gromov@dm.komisc.ru }
\maketitle

\begin{center}
Abstract
\end{center}

\noindent
More then forty years ago R.I.~Pimenov  introduced a new geometry --- semi-Riemannian one 
 --- as a set of geometrical objects consistent with a fibering 
$ pr: M_n \rightarrow M_m.$ He  suggested the heuristic principle according to which 
the physically different quantities (meter, second, coulomb etc.) are geometrically modelled  as space coordinates that are not superposed  by automorphisms. As  there is only one type of coordinates
in Riemannian geometry and only three types of coordinates in pseudo-Riemannian one,  a multiple fibered semi-Riemannian geometry is the most appropriate one for the treatment  of more then three different physical quantities as unified geometrical field theory.

Semi-Euclidean geometry $^{3}R_5^4$ with 1-dimensional fiber $x^5$ and 4-dimensional Minkowski space-time as a base is naturally interpreted as   classical electrodynamics. Semi-Riemannian geometry $^{3}V_5^4$ with the general relativity  pseudo-Riemannian space-time $^{3}V^4,$ and  1-dimen\-sio\-nal fiber $x^5,$ responsible for the electromagnetism, provides the unified field theory of gravitation and electromagnetism.

Unlike Kaluza-Klein theories, where the 5-th coordinate  appears in nondegenerate Riemannian or pseudo-Riemannian geometry, the theory  based on semi-Riemannian geometry is free from  defects of the former. In particular, scalar field does not arise.


\bigskip

\section{Introduction}  
The old problem of geometrical unification of gravity and electromagnetism goes back to the Kaluza-Klein theory \cite{K-21,K-26},
where electromagnetism is described by curvature in an extra spacelike dimension, i.\,e.  the fifth coordinate is introduced in pseudo-Riemanian geometry. However the physical dimension of electromagnetism is different from the dimension of space,  further\-more the fifth dimension is not observed as part of our everyday lives. To overcome  last difficulty, the fifth dimension is supposed to be  cyclic with very small radius. 
In addition there is also restriction on the set of transformations under which the theory is invariant. Indeed rotation in a plane spanning  the fifth coordinate $x^5$ and some space coordinate $x$ 
$$
x'=x\cos\varphi +x^5\sin\varphi, \quad
x'^5=x^5\cos\varphi -x\sin\varphi,
$$
immediately leads to the dependence of space-time coordinates on the fifth dimension what is obviously unsatisfactory.
Moreover an additional scalar field  arises in Kaluza-Klein type theories \cite{R-56,Ch-85}, 
which makes  Coulomb potential  everywhere regular.

Nevertheless the problem of unification gravitation with electromagnetism in the framework of one geometry is still actual.
This may help to incorporate some other interactions (weak interaction is the first candidate) in one geometrical picture. To overcome the disadventages of Kaluza-Klein theory the fibered semi-Riemannian geometry with degenerate metrics need be used.

The purpose of this paper is to explain the basic notions of semi-Riemannian geometry and to demonstrate on the example of Pimenov theory of gravitation and electromagnetizm how different physical quantities may be unified with the help of this geometry.

More then forty years ago   R.I.~Pimenov suggested \cite{P-64-1}--\cite{P-91} a unified theory of gravitation and electromagnetism where both fundamental interactions are naturally incorporated into the five dimensional semi-Riemanian geometry with degenerate metric.
This theory is free from disadvantages of Kaluza-Klein type theories. Mathematical definition  of multi-fibered semi-Riemanian geometry of arbitrary dimension was given by R.I.~Pimenov \cite{P-64-2}--\cite{P-68}.

 5-dimensional semi-Riemanian geometry can be formulated in a traditional way with the help of a real geometrical objects such as coordinates,
components of metric tensor etc. as well as can be obtained from Riemanian geometry with the help of nilpotent fifth coordinate \cite{P-91}.

Recently \cite{S-04} gravitation and electromagnetism have been  unified in the five dimensional space with degenerate metric and nilpotent fifth coordinate. But the scalar field is presented in this theory and Coulomb potential is regular at small distance.

\section{Geometrical modelling of a physical quantities } 

Geometrical modelling of  physical quantities is understood as unification of a different physical quantities (meter, second, coulomb etc.) 
 within the framework of  one  geometry. 
  Let us start with the simplest case of 2-dimensional space.

\begin{center}
\begin{figure}[htb]
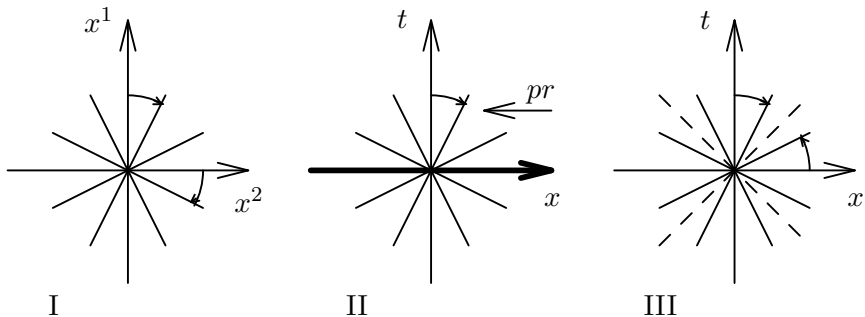

\hspace{4mm}  
\begin{texdraw}
\drawdim mm
\arrowheadtype t:V
\move(-16 0)\ravec(32 0)
\move(0 -15)\ravec(0 35)
\move(-5 -10)\lvec(5 10)
\move(-10 -5)\lvec(10 5)
\move(-5 10)\lvec(5 -10)
\move(-10 5)\lvec(10 -5)
\textref h:C v:C \htext(-4 20) {$x^1$}
\textref h:C v:C \htext(16 -4) {$x^2$}
\textref h:C v:C \htext(-10 -18) {I}
\move(0 0) \larc r:10 sd:-28 ed:0
\arrowheadsize l:1 w:1
\move(3.5 9.3) \ravec (1 -0.5)
\move(0 0) \larc r:10 sd:62 ed:90
\arrowheadsize l:1 w:1
\move(9.1 -3.7) \ravec (-0.7 -0.5)
\end{texdraw}
\hspace{6mm}  
\begin{texdraw}
\drawdim mm
\arrowheadtype t:V
\linewd 0.8 \move(-16 0)\ravec(32 0)\linewd 0.25
\move(0 -15)\ravec(0 35)
\move(-5 -10)\lvec(5 10)
\move(-10 -5)\lvec(10 5)
\move(-5 10)\lvec(5 -10)
\move(-10 5)\lvec(10 -5)
\move(16 8) \ravec(-9 0)
\textref h:C v:C \htext(14.5 10) {$pr$}
\textref h:C v:C \htext(-4 20) {$t$}
\textref h:C v:C \htext(16 -4) {$x$}
\textref h:C v:C \htext(-10 -18) {II}
\move(0 0) \larc r:10 sd:62 ed:90
\arrowheadsize l:1 w:1
\move(3.5 9.3) \ravec (1 -0.5)
\end{texdraw}
\hspace{6mm}  
\begin{texdraw}
\drawdim mm
\arrowheadtype t:V
\move(-16 0)\ravec(32 0)
\move(0 -15)\ravec(0 35)
\move(-5 -10)\lvec(5 10)
\move(-10 -5)\lvec(10 5)
\move(-5 10)\lvec(5 -10)
\move(-10 5)\lvec(10 -5)
\textref h:C v:C \htext(-4 20) {$t$}
\textref h:C v:C \htext(16 -4) {$x$}
\textref h:C v:C \htext(-10 -18) {III}
\move(0 0) \larc r:10 sd:0 ed:28
\arrowheadsize l:1 w:1
\move(3.5 9.3) \ravec (1 -0.5)
\move(0 0) \larc r:10 sd:62 ed:90
\arrowheadsize l:1 w:1
\move(9.3 3.7) \ravec (-0.5 0.6)
\lpatt(2 2) \move(-10 -10)\lvec(10 10) 
\move(-10 10)\lvec(10 -10) 
\end{texdraw}
\caption{ A bundle  of lines in Euclidean  (I), semi-Euclidean  (II) and pseudo-Euclidean (III) planes.}
\label{fig1}
\end{figure}
\end{center}

For a bundle  of lines through a point in a plane one of the three possibilities is realized  \cite{P-65-3} (see  Fig. \ref{fig1}):

I. Any line of the bundle is postulated.

II. There is one non postulated (isolated) line in the bundle.

III. There are two or more non postulated lines in the bundle.

The property of line to be postulated is conserved by the plane automorphisms.

Postulate I leads to the Euclidean geometry in the plane. Any two lines of the bundle are superposed  by rotations  around the point.
This means that only one physical quantity can be modelled by Euclidean geometry. 
The same is true for n-dimensional  Euclidean space.
Very often lines of Euclidean geometry are  interpreted  as a space length, i.\,e. $[x^1]=\ldots=[x^n]=[\mbox{length}].$ 
 
Postulate III leads to the pseudo-Euclidean geometry in the plane, where there are three types of lines: with positive, negative and zero length.
These types of lines are not  transformed to each other by rotations  around the point (Lorentz transformations).
Therefore only three different physical quantities can be incorporated in pseudo-Euclidean  geometry. For kinematical interpretation these are time-like, space-like and light-like lines. The same is true for n-dimensional pseudo-Riemannian  space. 

Postulate II leads to the fibered semi-Euclidean (or Galilean) geometry in the plane which is defined by the projection $pr$ with one dimensional base $\left\{t\right\}$ and one dimensional fiber $\left\{x\right\}.$ This plane is interpreted as  classical 
(1+1) kinematics  with absolute time $t$ and absolute space $x$. Rotations (or Galilei boosts) superpose any two time-like lines but don't 
superpose time-like lines with space-like line. 
Two different physical quantities (space and time) are modelled by two types of lines on the semi-Euclidean  plane.

Semi-Riemanian geometry is a fibered geometry with degenerate metric. 
If the base of semi-Riemanian space  has Euclidean geometry then the number of line types remains the same under the adding of extra dimensions to the base.
If it has pseudo-Riemanian geometry then only one additional type of line with nonzero length   appears. 
This exhausts all posibilities for incorporating new physical quantities.

Different situation takes place  when  adding  extra dimensions to the fiber. 
Riemanian, pseudo-Riemanian   or  semi-Riemanian geometry can be realized in the  fiber. For example, if the fiber $\left\{x,y\right\}$          has semi-Riemanian geometry then there is projection $pr_2$ with the base $\left\{x\right\}$ and the new fiber $\left\{y\right\}.$               So the third type of line  appears in this geometry and it   gives an opportunity  for modelling some third physical quantity                 ${[y]}\neq [t]\neq {[x]}.$

\begin{center}
\begin{figure}[htb]
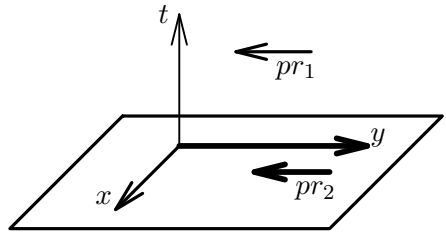

\hspace{25mm}
\begin{texdraw}
\drawdim mm 
\setunitscale 0,5
\arrowheadtype t:V
\move(0 0)\ravec(0 35)
\textref h:C v:C \htext(-4 35) {$t$}
\textref h:C v:C \htext(53 2) {$y$}
\textref h:C v:C \htext(-20 -14) {$x$}
\textref h:C v:C \htext(35 -12) { $pr_2$}
\textref h:C v:C \htext(30 20) { $pr_1$}
\linewd 0.8 \move(0 0)\ravec(-17 -17)
\move(35 25)\ravec(-20 0)
\move(-15 8)\lvec(70 8)\lvec(40 -22)\lvec(-45 -22)
\lvec(-15 8)
\linewd 1.6 \move(0 0)\ravec(50 0)
\move(40 -7)\ravec(-20 0)
\end{texdraw}
\caption{ Types of lines in twice fibered space. } 
\label{fig1-1}
\end{figure}
\end{center}

Similar algorithm for construction of multi-fibered spaces can be applied  without limit.
In other words, multi-fibered geometry with the subsequently enclosed  projections can be successfully used for   incorporating
of any number of physical quantities in one geometry.

The above examples  illustrats  the {\it heuristic principle} \cite{P-91}:
{\bf   quantities with different physical  dimensions   cannot be geometrically superposed  by automorphisms.  }

\section{Semi-Riemannian geometry $ ^{3}V^4_5$ as space-time\- -electromagnetism}   

\subsection{Definition of semi-Riemannian space  $ ^{3}V^4_5$}

The semi-Riemannian space $ ^{3}V^4_5$ is described by the projection
$pr:\; ^{3}V^4_5 \rightarrow ^{3}V_4 $
with 4-dimensional base $ ^{3}V_4$ and one-dimensional fiber $\left\{x^5\right\}.$
Pseudo-Riemannian geometry with the nondegenerate metrics  $g_{\mu\nu},\; \mu,\nu=1,2,3,4$ of signature $(+---)$ is set in the base.
It follows from  consistency with fibering that components $g_{\mu\nu}$  depend on base coordinate $x^1,\ldots,x^4.$
Metrics in the fiber is defined by the component $g_{55}\neq 0 $  depending on all coordinates $x^1,\ldots,x^5.$
This component can be regarded as a scale for fifth coordinate.
Remaining components $g_{\mu 5}(x^1,\ldots,x^5)=g_{5\mu}(x^1,\ldots,x^5)$  define the 4-distribution
\begin{equation}
\omega = \left\{g_{\alpha 5}dx^{\alpha}=0\right\},
\label{2.2-1}
\end{equation}
which is transversal to the fiber.
As a result the metric flagtensor of the semi-Riemannian space $ ^{3}V^4_5$ has the form
$g_{\alpha\beta}=\left(g_{\mu\nu},g_{\mu 5 },g_{55} \right). $ 

Coordinate transformations  consistent with the fibering  can be written  as
\begin{equation}
\left\{
\begin{array}{rcl}
x^{\mu'}&=&f^{\mu'}(x^1,\ldots ,x^4)  \\
x^{5'}&=&f^{5'}(x^1,\ldots ,x^4,x^5) 
\end{array}
\right.
\label{2.2}
\end{equation}
\begin{equation}
(D^{\alpha'}_\beta)=\left(\begin{array}{cc}
(D^{\mu'}_\nu)& 0\\
(D^{5'}_\nu) & D^{5'}_5\end{array}\right), 
  \quad \det (D^{\mu'}_\nu) \neq 0, \,\, D^{5'}_5\neq 0,
\label{2.3}
\end{equation}
where ${\cal D}_{\beta}^{\alpha'}=\partial_{\beta}x^{\alpha'}. $
The metric flagtensor $g_{\alpha\beta}=\left(g_{\mu\nu},g_{\mu 5 },g_{55} \right)$ is transformed under 
 (\ref{2.2}),(\ref{2.3}) by the formulas
$$ 	
g_{\mu'\nu'}=g_{\mu\nu}{\cal D}^\mu_{\mu'}{\cal D}^\nu_{\nu'}, \quad		
g_{5'5'}=g_{55}{\cal D}^5_{5'}{\cal D}^5_{5'}, 
$$
\begin{equation}
g_{5'\mu'}=g_{5\mu}{\cal D}^5_{5'}{\cal D}^\mu_{\mu'} + g_{55}{\cal D}^5_{5'}{\cal D}^5_{\mu'}.		
\label{2.4}
\end{equation}
Here and below  summation on repeating  indices to be  understood.
It is always possible to make $g_{55}=1$ by the scale changing
${\cal D}^5_{5'}=\frac{\partial x^5}{\partial x^{5'}}=\frac{1}{\sqrt{g_{55}}} $;
assuming this, we have (${\cal D}^5_{5'}={\cal D}^{5'}_{5}=1$)
\begin{equation}		
g_{5'\mu'}=g_{5\mu}{\cal D}^5_{5'}{\cal D}^\mu_{\mu'} + g_{55}{\cal D}^5_{5'}{\cal D}^5_{\mu'} = g_{5\mu}{\cal D}^\mu_{\mu'} + {\cal D}^5_{\mu'}.		
\label{2.6}
\end{equation}
If one  assumes that in the base, which is interpreted as space-time, nothing happens, i.\,e. ${\cal D}^\mu_{\mu'}={\delta}^\mu_{\mu'}, $
then the transformations of metric flagtensor components $g_{5\mu}=g_{\mu 5}$ look like  adding  arbitrary function gradients
\begin{equation}		
g_{5\mu} \longmapsto g_{5\mu} + \partial_{\mu}x^5, \quad g_{\mu 5} \longmapsto g_{\mu 5}+ \partial_{\mu}x^5.
\label{2.7}
\end{equation}

Invariants of semi-Riemannian geometry $ ^{3}V^4_5$ are 
pseudo-Riemannian length in the base
\begin{equation}		
ds=\sqrt{g_{\mu\nu}dx^\mu dx^\nu},
\label{2.8}
\end{equation}
and Euclidean length in the fiber
\begin{equation}		
ds_{(2)}=\sqrt{g_{55}dx^5 dx^5}=|dx^5|\quad (\mbox{with}\; g_{55}=1 ).
\label{2.9}
\end{equation}
As far as the angle  between general position vector $d{\bf x}=(dx^\mu,dx^5)$ and  fiber vector $\delta{\bf x}=(0,\delta x^5)$  equals
infinity,   the supplemental angle $\psi$ is taken as the third invariant
\begin{equation}		
\psi=g_{\mu 5}\frac{dx^\mu}{ds} +\frac{dx^5}{ds}. 
\label{2.10}
\end{equation}
$\psi$ is the angle between  vector $d{\bf x}$ and its projection on the base in 2-plane formed by $d{\bf x}$ and $\delta{\bf x}.$ 

Thus, semi-Riemannian geometry $^3V^4_5$ is defined by the set of objects described above  with transformation properties (\ref{2.2})-(\ref{2.4}).
			
In order to avoid terminological misunderstanding let us stress that the fibering $pr$ has nothing to do with the principal bundle, where some group  acts on the fiber. In the last approach something similar to a tangent space $(dx^1,\ldots,dx^n)$ is built on the space $(x^1,\ldots,x^n)$
and then unified object $(x^1,\ldots,x^n,dx^1,\ldots,dx^n)$ is regarded as having base $(x^1,\ldots,x^n)$ and fiber $(dx^1,\ldots,dx^n).$
In semi-Riemannian geometry the fibering "`takes place"'  in the space $(x^1,\ldots,x^n)$ itself. 
The reader should not confuse the terminology and methods of these different notions of fibering.

\subsection{Interpretation of semi-Euclidean geometry  $^3R^4_5 $ as classical electrodynamics} 

The fiber can be regarded as an "`inner space"' of a particle. 
The fiber geometry in semi-Riemannian space is  completely defined by the metrics $g_{55}(x^1,\dots, x^5)$ and does not depend on the base metrics $g_{\mu\nu}(x^1,\dots, x^4)$ (which is  not the case in nondegenerate geometry).
The remaining components $g_{\mu 5} $  are functions of all coordinates $x^1,\dots, x^5.$

Let us impose on the space $^3V^4_5 $ an additional condition, namely,  {\bf the fiber has Euclidean geometry which does not depend on base coordinates $x^1,\dots, x^4$. }
Then in a neighbourhood (instead of just one point) there exists such  coordinate system that all components of flagaffine connection are equal to zero and $\partial_{\mu}g_{55}=0,\;\partial_{5}g_{55}=0,\;\partial_{5}g_{\mu 5}=0,$ which implies
$g_{55}=const=1,$ $g_{\mu 5}=g_{\mu 5}(x^1,\ldots,x^4).$

With this additional condition  the semi-Riemannian space of zero curvature (or semi-Euclidean space) can be defined.
The base of semi-Euclidean space is Minkowskian space-time with coordinates $x^1=t,\, x^2=x,\, x^3=y,\, x^4=z$ and metric tensor
$g_{\mu\nu}=\mbox{diag}(1,-1,-1,-1,).$
The fiber is Euclidean line $\{x^5\}$ and $g_{55}=1.$
Besides the metric flagtensor has nonzero components $g_{\mu 5}(t,x,y,z).$
Invariants (\ref{2.8}),(\ref{2.9}) can be written in integral form
\begin{eqnarray} 
s&=&\sqrt{t^2-x^2-y^2-z^2}, \quad t \geq \sqrt{x^2+y^2+z^2}, \nonumber \\
s_{(2)}&=&|x^5|, \quad t=x=y=z=0. 
\label{2.11}
\end{eqnarray}	
Coordinate transformations consistent with the fibering  are as follows:
\begin{equation} 
\left\{  \begin{array}{rcl}
(\tilde t,\tilde x,\tilde y,\tilde z)&=& {\cal P}(t,x,y,z),     \\
\tilde x^{5}&=& x^5 +  f(t,x,y,z), 
\end{array}
\right.
\label{2.13}
\end{equation}	
where ${\cal P}$ is an element of the Poincare group and $f(t,x,y,z) $ is an arbitrary function.

The action of classical electrodynamics is invariant under Poincare (Lorentz) group and is defined up to adding 
the arbitrary function of space-time coordinates. It follows from  (\ref{2.13}) that the fifth coordinate $x^5$ has  similar property and therefore can be interpreted as the action.

The 4-dimensional vector-potential $A_\mu$ of electromagnetic field is defined up to adding the gradient $\partial_\mu f $
of an arbitrary function of space-time coordinates. It follows from (\ref{2.7}) that the components $g_{\mu 5}$ of metric tensor
have just similar properties and therefore can be identified with vector-potential $A_\mu=g_{\mu 5}.$
Maxwellian field stress tensor   is obtained in a standard way
$F_{\mu\nu}=\partial_\mu A_\nu - \partial_\nu A_\mu.$

The action must be invariant under the automorphism group. Let us form it from the invariants (\ref{2.8}),(\ref{2.10}) of semi-Euclidean space
$^3R_5^4$
\begin{equation}
S= a\int_A^B ds + b\int_A^B \psi ds 
=S_p + S_{int}, 
\label{2.16}
\end{equation}
where $a,b$ are dimension factors. The third invariant $\int dx^5=x^5(x^\mu)$ (\ref{2.9}) is  an arbitrary function of space-time coordinates, therefore it can be omitted. For $a=-mc,$ where $m$ is the particle mass, $c$ the  light velocity we have
$$ 
S_p= -mc\int_A^B ds = -mc\int_A^B \sqrt{dt^2-dx^2-dy^2-dz^2} = 
$$
\begin{equation}
= -mc^2\int_{t_1}^{t_2} \sqrt{1-\frac{v^2}{c^2}}dt,
\label{2.16-1}
\end{equation}
i.\,e. the action for a free particle \cite{LL-88}, \S 8.	
Taking (\ref{2.10}) into account  we have $\int \psi ds=\int A_\mu dx^\mu + \int dx^5.$ The second term can be omitted, and for 
$b=-\frac{e}{c},$ where $e$ is the electron charge, we  obtain
\begin{equation}
S_{int}= -\frac{e}{c}\int_A^B A_\mu dx^\mu,
\label{2.17}
\end{equation}	
i.\,e. the action which describes the interaction of charged particle moving in a given electromagnetic field with this field
 \cite{LL-88},  \S 16.	
	
There is no  field action	among  geometrical invariants of the semi-Euclidean space  $^3R^4_5. $   However there is the electromagnetic field stress tensor $F_{\mu\nu}.$
Therefore, introducing a new postulate:  {\bf the field action must be an invariant function of electromagnetic field stress tensor}, we obtain in a standard way \cite{LL-88},  \S 27
\begin{equation}
S_{f}= -\frac{1}{16\pi}\int F_{\mu\nu}F^{\mu\nu} d^4x.
\label{2.18}
\end{equation}
Sum of the terms 	(\ref{2.16-1})--(\ref{2.18})
\begin{equation}
S=S_p + S_{int} + S_{f}
\label{2.19}
\end{equation}
gives the complete action for a particle in an electromagnetic field.

\subsection{ $^3V^4_5 $ as space-time-electromagnetism}

In semi-Riemannian space $^3V^4_5 $ for invariantly selected coordinate system, where $g_{55}=1$, the metric components
 $g_{\mu 5}(x)=A_\mu (x), \;x=(x^1,x^2,x^3,x^4)$ are interpreted as 4-vector-potential  of electromagnetic field.
The pseudo-Riemannian metrics of the general relativity space $^3V_4$ is specified by the components $g_{\mu\nu}(x).$
The fiber geometry does not affect  the geometry of the base  $^3V_4.$

To obtain gravitation field equations in the space $^3V_4$ its scalar curvature $R$ is used instead of invariant (\ref{2.8}). The variation of sum of the action for gravitation field \cite{LL-88},  \S 93
\begin{equation}
S_{g}= -\frac{1}{16\pi G}\int R\sqrt{-g} d^4x,
\label{2.20-a}
\end{equation}
where $G$ is gravitation constant, $g=\mbox{det}(g_{\mu\nu}),$ and the action (\ref{2.18}) for electromagnetic field gives in result
Einstein equations 
\begin{equation}
R_{\mu\nu} - \frac{1}{2}Rg_{\mu\nu} +\Lambda g_{\mu\nu} = -8\pi G T_{\mu\nu},
\label{2.20}
\end{equation}
where  $R_{\mu\nu}=R_{\mu\rho\nu}^{\rho}$ is Ricci tensor, curvature tensor
$R_{\mu\rho\nu}^{\pi}=\partial_{\rho}\Gamma_{\nu\mu}^{\pi} - \partial_{\nu}\Gamma_{\rho\mu}^{\pi}
+ \Gamma_{\rho\tau}^{\pi}\Gamma_{\nu\mu}^{\tau} - \Gamma_{\nu\tau}^{\pi}\Gamma_{\rho\mu}^{\tau}$
is expressed by Christoffel symbols
$\Gamma_{\mu\nu}^{\pi}= \frac{1}{2}g^{\pi\tau}\left(\partial_{\nu}g_{\tau\mu} + \partial_{\mu}g_{\tau\nu} - \partial_{\tau}g_{\mu\nu}\right),$
 $\;g^{\mu\tau}g_{\tau\nu}=\delta^\mu_\nu,  $
$R=R^\mu_\mu=g^{\mu\nu}R_{\nu\mu}\;$ is scalar curvature,    $\Lambda$ is cosmological constant,  $T_{\mu\nu}$ is energy-momentum tensor.
Thus left hand side of Einstein equations is  determined by the metrics $g_{\mu\nu}(x)$ and their right hand side  is  determined by the energy-momentum tensor of the system under consideration
\begin{equation}
T_{\mu\nu}=g_{\mu\rho}g_{\nu\tau}T^{\rho\tau}, 
\label{2.21}
\end{equation}
where 
\begin{equation}
 T^{\rho\tau}= \frac{1}{4\pi}\left(-F^{\rho\nu}F^{\tau}_{\nu} + \frac{1}{4}g^{\rho\tau}F_{\mu\nu}F^{\mu\nu}  \right)
\label{2.22}
\end{equation}
is the energy-momentum tensor of electromagnetic field without charges.

So the space $^3V^4_5 $ gives rise to the standard equations for interactions of gravitational and electromagnetic fields demonstrating that their are incorporated in the framework of semi-Riemannian geometry.

\subsection{Comparison with the Kaluza-Klein theories  }

The unification of general relativity and electromagnetism within the framework of 5-dimensional geometry goes back to the papers  \cite{K-21,K-26}.  
In Kaluza-Klein type theories 4-dimensional pseudo-Riemannian space-time of general relativity is extended to 5-dimensional one by adding on  extra dimension responsible for electromagnetism \cite{R-56,Ch-85,KP-00}. 
Since  the fifth coordinate is introduced  within the framework of pseudo-Riemannian geometry  it can be superposed with any space-time coordinate by an automorphism and therefore cannot model a new physical quantity. To avoid this difficulty the fifth dimension is supposed to be "`curled up small"' so that it cannot be directly considered as ordinary space or time dimension.
Moreover, additional scalar field $\chi$ appears as $g_{55}=1+\chi$  in Kaluza-Klein type models. This scalar field modifies electromagnetic  potential  in such a way that it is everywhere  regular and is close to the Coulomb  potential at all except very small length scales.

The use of semi-Riemannian geometry $^{3}V^4_5$ allows  to overcome  above mentioned difficulties, namely:

1. The independence of $g_{\mu\nu}$ of $x^5$ follows from the geometrical axioms. It is possible, however  not necessary, to suppose the fifth coordinate to be cyclic with very small radius.

2.  The independence of $g_{\mu 5}$ of $x^5$ follows from the additional assumption that
the fiber has Euclidean geometry which does not depend on base coordinates.

3. No scalar fields arise. 

4. Space-time  and  the fiber geometries can be taken absolutely independent of each other.

5. Einstein equations in space-time impose no  restrictions on  extra coordinates.

6. Gradient invariance of electromagnetic vector-potential is automatically  a consequence of transformations of metric flagtensor components.

7. The unified theory can be applied to the empty space-time of special relativity, where  electromagnetic field is present.  
(This is not possible in pseudo-Riemannian case when $g_{55}=1,\;\partial_5 g_{\mu 5}=0$ because of the formula
$2R_{\mu 5\mu 5}=2\partial_\mu \partial_5 g_{\mu 5} - \partial_5\partial_5 g_{\mu\mu} - \partial_\mu \partial_\mu g_{55},$ 
which  is not the case in semi-Riemannian geometry.)
The action for charged particle interacting with a given electromagnetic field is the sum of invariants of semi-Euclidean geometry.
 
8. The heuristic principle  demanding that different physical quantities should not be geometrically superposed  by automorphisms is fulfilled.


\section{Semi-Riemannian space  $V^4_5$ with nilpotent fifth coordinate } 

In the previous sections semi-Riemannian geometry was constructed for  real objects as the  structure consistent with fiberig.
At the same time geometry with degenerate metrics can be realized with the help of nilpotent objects.
In particular, fiber space can be obtained from Riemannian space by multiplying  fiber coordinates by nilpotent unit $\iota, \;\iota^2=0.$ 
Similar approach had been  useful for investigating  group contractions \cite{G-90}.

Let us obtain $V^4_5$ from  Riemannian space $V_5$ by substituting $\iota x^5$ instead of $x^5.$
We shall demand that the following heuristic rules be fulfiled:
for a real $a$ the expression $\frac{a}{\iota}$ is defined only for $a=0,$ however  $\frac{\iota}{\iota}=1;$
furthermore
\begin{equation}
 \sqrt{a^2 + \iota^2b^2}= \left\{ 
\begin{array}{cc} 
|a|,& \mbox{if} \; a\neq 0, \nonumber \\
\iota |b|, &	\mbox{if} \; a=0. \nonumber
\end{array}
 \right.  
\label{3.1}
\end{equation} 

Substitution $x^5  \rightarrow \iota x^5$ induces another  one  $g_{\mu 5}\rightarrow \iota g_{\mu 5},$ i.\,e. metric tensor has nilpotent components
\begin{equation}
g_{\alpha\beta}=\left(\begin{array}{cc}
g_{\mu\nu} & \iota g_{\mu 5}\\
\iota g_{\mu 5} & g_{55} \end{array}\right).
\label{3.2}
\end{equation} 
The  nondegeneracy condition $\det (g_{\alpha\beta})=\det (g_{\mu\nu})g_{55}\neq 0$ implies  nondegeneracy  in the base 
$\det (g_{\mu\nu})\neq 0$ and in the fiber $g_{55}\neq 0.$ Inverse tensor $g^{\alpha\beta}$ is easily obtained from equations $g^{\alpha\gamma}g_{\gamma\beta}=\delta^\alpha_\beta$ and is as follows
\begin{equation}
g^{\alpha\beta}=\left(\begin{array}{cc}
g^{\mu\nu} & -\iota g_{55}^{-1}g^{\mu\nu} g_{\nu 5}\\
-\iota g_{55}^{-1} g_{\nu 5}g^{\nu\mu} & g_{55}^{-1} \end{array}\right),
\label{3.3}
\end{equation} 
where $g^{\mu\rho}g_{\rho\nu}=\delta^\mu_\nu.$ For  $g_{\mu\nu}=\mbox{diag}(1,-1,-1,-1)=g^{\mu\nu},\;g_{55}=1, $ we have  $g^{15}=-g_{15},\;g^{k5}=g_{k5},\;k=2,3,4. $

The entries   $D^{\mu'}_5=\frac{\partial x^{\mu'}}{\partial x^{5}}$ of transformation matrix are substituted for
$D^{\mu'}_5=\frac{1}{\iota}\frac{\partial x^{\mu'}}{\partial x^{5}},$ i.\,e. 	
$\frac{\partial x^{\mu'}}{\partial x^{5}}=0,$
and entries   $D_{\mu}^{5'}=\frac{\partial x^{5'}}{\partial x^{\mu}}$ are substituted for
$\iota\frac{\partial x^{5'}}{\partial x^{\mu}}=\iota D_{\mu}^{5'}.$
Therefore differentials of coordinate functions are transformed as
\begin{equation}
\left(\begin{array}{c}
(dx^{\mu'})\\
\iota dx^{5'}
\end{array}\right)
=\left(\begin{array}{cc}
(D^{\mu'}_\mu)& 0\\
\iota(D^{5'}_\mu) & D^{5'}_5\end{array}\right) 
\left(\begin{array}{c}
(dx^{\mu})\\
\iota dx^{5}
\end{array}\right),  
\label{3.3-1}
\end{equation}
and coordinate transformations are  consistent with fibering form (\ref{2.2}),(\ref{2.3}). 
From general formula $g_{\alpha'\beta'} = g_{\alpha\beta}D^\alpha_{\alpha'}D^\beta_{\beta'} $ 
for the metrics (\ref{3.2}) with the help of matrix (\ref{3.3-1}) we obtain the transformations (\ref{2.4})   of metric tensor. 

Invariants of semi-Riemannian space $V^4_5$ are obtained from those of Riemannian space $V_5,$ namely
$$  
 ds=\sqrt{g_{\alpha\beta}dx^\alpha dx^\beta} = \sqrt{g_{\mu\nu}dx^\mu dx^\nu + 
 \iota^2\left(g_{55}dx^5dx^5 + 2g_{\mu 5} dx^\mu dx^5 \right)} = 
$$
\begin{equation}	 
= \left\{
\begin{array}{cc}
\sqrt{g_{\mu\nu}dx^\mu dx^\nu}=ds, & \mbox{if}\;\;\exists \mu \;\; dx^\mu \neq 0 - \mbox{ length in the base}, \\
\iota |dx^5| = \iota ds_{(2)}, & \mbox{if }\;\;\forall \mu \;\; dx^\mu =0 - \mbox{length in the fiber}.
\end{array} 
\right.
\label{3.8}
		\end{equation}
Supplementary angle between general position vector $d{\bf x}=(dx^\mu,\iota dx^5) $ and fiber vector $\delta{\bf x}=(0,\iota \delta x^5)$		
is deduced from the formula	for sine of supplementary  angle   $\psi$ between  vectors $d{\bf x}$ and $\delta {\bf x}$ in Riemannian space
\begin{equation}	
\cos(\frac{\pi}{2}- \psi)=\sin \psi = \frac{g_{\alpha\beta}dx^\alpha \delta x^\beta}{|d{\bf x}||\delta {\bf x}|}
\label{3.9}
		\end{equation}
by  substitutions  $x^5 \rightarrow \iota x^5,\;\; g_{\mu 5} \rightarrow \iota g_{\mu 5},\;\; \psi \rightarrow \iota \psi.$	
As far as  $|d{\bf x}|=\sqrt{g_{\mu\nu}dx^\mu dx^\nu}=ds,\; |\delta {\bf x}|=\iota \delta x^5,\; \sin \iota \psi =\iota \psi$ 
(functions of nilpotent arguments are defined by their Taylor expansion ) and
$g_{\alpha\beta}dx^\alpha \delta x^\beta = $   $\iota g_{\alpha 5}dx^\alpha \delta x^5 =$ 
$\iota \delta x^5 \left(g_{55}\iota dx^5 + \right.$    $\left.    \iota g_{\mu 5}dx^\mu \right) = $
$  \iota^2 \delta x^5 \left(g_{\mu 5}dx^\mu + dx^5\right),$
then from (\ref{3.9}) we have
\begin{equation}	
\iota  \psi = \frac{\iota^2 \delta x^5 \left(g_{\mu 5}dx^\mu + dx^5\right)}{ds \iota \delta x^5} =
\iota\left(g_{\mu 5}\frac{dx^\mu}{ds} + \frac{dx^5}{ds}\right),
\label{3.10}
		\end{equation}
what  coincides with 	(\ref{2.10}) after 		cancelling $\iota$.	

The fact that the base geometry does not depend on  fiber geometry comes  automatically. 
Indeed, derivative
$
\frac{1}{\iota}\frac{\partial g_{\mu\nu}}{\partial x^5}
$	
is defined only for $\frac{\partial g_{\mu\nu}}{\partial x^5}=0, $ i.\,e. $g_{\mu\nu}(x^1,\ldots,x^4)$ does not depend on $x^5.$

In a similar way pseudo-Riemannian geometry $^3V_4$ in the base can be obtained from Riemannian one $V_4$ by "`analytic continuation"' of locally orthogonal coordinates  substituting $ix^k$ for  $x^k, \; k=2,3,4.$

Let us note that recently the interest was shown to the problem of unifying gravitation with electromagnetism in a space with nilpotent fifth coordinate  \cite{S-04}. However semi-Riemannian geometry was not constructed there, therefore the scalar field is present in this theory and electromagnetic potential is  everywhere regular.

\section{Conclusions}  

R.I.~Pimenov's unified geometrical theory of gravitation and electromagnetism is based on the  multi-fibered semi-Riemannian geometry developed by himself. This theory is formulated taking into account the heuristic principle according to which physically different quantities cannot be superposed by automorphisms. 
   In other words, for unification of gravitation (or space-time) with some other fundamental interactions (electromagnetic or possible weak) it is necessary to use $D$-dimensional fiber space ($D>4$) with degenerate metrics. The base of this space is 4-dimansional pseudo-Riemannian space-time of general relativity and extra dimensions, which are responcible for other interactions, belong to the fiber.
The particular  case of single-fibered semi-Euclidean space $^{3}R_5^4$ with one-dimensional fiber $x^5$ 
and 4-dimensional Minkowski space-time as the base  is naturally interpreted as classical electrodynamics. 
Both fundamental gravitational and electromagnetic interactions are incorporated in one
semi-Riemannian geometry $^{3}V_5^4$ with  general relativity space-time as the base of the fibering and  one-dimensional fiber $x^5$ responsible for electromagnetism.

Unlike   Kaluza-Klein type theories, where  fifth dimension  appears in the context of nondegenerate Riemannian or pseudo-Riemannian geometry,  Pimenov theory based on semi-Riemannian geometry does not have restrictions on   admissible transformations of 5-dimensional space  as well as additional scalar field
 which modifies Coulomb potential at a small distances. The last property is not compatible with experimental data.
 
5-dimensional semi-Riemannian geometry is formulated in real geometrical notions  (coordinates, metric tensor components etc.)
but can also be obtained from  Riemannian geometry by using   nilpotent fifth coordinate.  

\section*{Acnowledgments}

This work has been supported in part by the Russian Foundation for Basic Research, grant 08-01-90010-Bel-a.

\end{document}